\documentclass[12pt]{iopart}
\newcommand{\be}{\begin{equation}}
\newcommand{\ee}{\end{equation}}
\newcommand{\bea}{\begin{eqnarray}}
\newcommand{\eea}{\end{eqnarray}}

\usepackage{graphicx}% Include figure files
\usepackage{bm}% bold math
\usepackage{iopams}
\usepackage{setstack}
\begin{document}

\title{Landau quantization effects in ultracold atom-ion collisions}
\author{Andrea Simoni and Jean-Michel Launay}
\address{
Institut de Physique de Rennes,  UMR 6251 CNRS-Universit\'e de
Rennes 1, 35042 Rennes Cedex, France}

\begin{abstract}

We study ultracold atom-ion collisions in the presence of an external
magnetic field. At low collision energy the field can drastically modify
the translational motion of the ion, which follows quantized cyclotron
orbits. We present a rigorous theoretical approach for the calculation of
quantum scattering amplitudes in these conditions. Collisions in different
magnetic field regimes, identified by the size of the cyclotron radius
with respect to the range of the interaction potential, are investigated.
Our results are important in cases where use of a magnetic field to
control the atom-ion collision dynamics is envisioned.

\end{abstract}

\date{\today}
\maketitle
\section{Introduction}

The central role played by the collisional interaction in
determining the macroscopic properties of cold atomic gases has
motivated extensive theoretical and experimental studies of neutral
atom collisions at very low collision energies over the last decade.
For instance, magnetic Feshbach resonances have been discovered and
theoretically analyzed in several atomic species (see, {\it e.g.},
\cite{2000-HW-PRA-052704,2008-AS-PRA-052705}). Such resonances allow
the interatomic interaction to be finely tuned, and have been used
to explore a variety of phenomena such as the BEC-BCS crossover
\cite{2004-CAR-PRL-040403} and strongly correlated quantum phases
\cite{2006-TV-NP-692}.

While atom-ion collisions at standard temperatures are a well
established subject \cite{1981-JBD-RMP-287}, their behaviour at very low
collision energy has only recently begun to raise a rapidly increasing
interest. Atom-ion systems are in particular very promising candidates
for the realization of collisional quantum gates \cite{2010-HD-PRA-012708}
and for the study of ionic impurities in neutral Bose-Einstein condensates
\cite{2002-RC-PRL-093001}. Experimental progress makes now ultracold
atom-ion systems addressable in a controlled way \cite{2010-CZ-NAT-388}.
First measurements of elastic cross sections and sympathetic cooling
rates between a Bose condensed gas and a single ion held in a Paul trap
have been reported in Refs.~\cite{2010-CZ-NAT-388,2009-ATG-PRL-223201}.
Charge exchange has also been theoretically analyzed both for its
interest in fundamental molecular physics and in order to quantify
harmful sources of atom-loss or decoherence in the experiments
\cite{2009-ZI-PRA-010702,2003-OPM-PRA-042705}.

The presence of magnetically tuned Feshbach resonances induced
by the hyperfine interaction has been demonstrated in free space
in~\cite{2009-ZI-PRA-010702}. Such resonances could in principle be
engineered to increase the rapidity of collisional quantum gates or to
manipulate the many-body properties of a neutral gas in the presence
of a charged impurity \cite{2010-JG-PRA-041601}. However, non trivial
effects have to be taken into account to describe quantitatively atom-ion
collisions in most non idealized situations, such as trap confinement
\cite{2007-ZI-PRA-033409}. An additional feature neglected in previous
work \cite{2009-ZI-PRA-010702} is the influence of the magnetic field
on the translational motion of the ion, which will classically follow
helicoidal rather than rectilinear trajectories. Quantum mechanically,
motion in the plane transverse to the field has to be quantized,
resulting in the discretization of the continuum in Landau levels
\cite{BK-1999-LDL}. 

The dramatic influence of even a weak magnetic field on electron
photodetachment was observed a long time ago \cite{1978-WAMB-PRL-1320}
and modeled using a frame transformation approximation, which relies
on the much smaller length scale of the interatomic potential compared
to the cyclotron radius~\cite{1987-CHG-PRA-4236}. However, the frame
transformation approach cannot be applied to ultracold atom-ion collisions
since the typical length scales become comparable and cannot therefore
be separated already at commonly used field intensities of a few hundred
Gauss (see below). Numerically exact $R$-matrix calculations have been
carried out for hydrogen photoionization but were limited to strong
magnetic fields~\cite{1991-QW-PRA-7448}. The main goal of our paper
is to develop a rigorous theory applicable to atom-ion collisions in a
magnetic fields of arbitrary strength.

The paper is organized as follows. The system and general scattering
properties in a magnetic field are introduced in Sec.~2.  Sec.~3
presents the numerical methods. In Sec.~4 our framework is applied to
the scattering of Ca$^+$ ions and Na atoms. A variety of threshold,
quantum interference and resonance phenomena are identified and
discussed. Concluding remarks and perspectives end this work.

\section{Theoretical model}

\subsection{Background}

We consider a low energy collision between a neutral atom with a singly
charged ion in the presence of a uniform static magnetic field $\vec B$
oriented in the $z$ direction. We suppose that both particles have
spherical symmetry, vanishing internal spin, and that the atom
has an infinite mass. The latter assumption allows the number of spatial
degrees of freedom to be reduced from six to three, thus avoiding
complications related to the separation of the center of mass
motion \cite{1977-JEV-ANNP-431}.

The spherically symmetric interparticle molecular potential $V(r)$ has a
long-range behaviour determined by the second order charge induced-dipole
interaction $-C_4 r^{-4}$, where $2C_4$ is the static dipolar polarizability of the
neutral atom. The detailed shape of the potential well is immaterial
for ultracold collisions and it can be to excellent approximation
described by energy-independent quantum defects or scattering
lengths \cite{2007-ZI-PRA-033409}.  For this work we take therefore
a simple analytical form $V(r)=D_e \left(x^{-8}-2 x^{-4} \right)$
with $x=r/r_e$. The potential parameters $D_e$ and $r_e$ are adjusted
to approximately agree with the published CaNa$^+$ potentials and to
reproduce the $C_4=81.3$~a.u. value \cite{2003-OPM-PRA-042705}. The
$s$-wave scattering length is currently unknown for any atom-ion
system. We perform therefore some exploratory studies for different scattering
lengths by adding a small correction to the bottom of the potential well
to vary the value of $a_s$ in the numerical calculations~\cite{1996-ET-JRNIST-505}.

The Hamiltonian for a singly positively charged ion is 
\be
H=\frac{\left( \vec{p} - e \vec{A} \right) ^2}{2m}+V
\ee
where $\vec A$ is the vector potential associated to $\vec B$, $\vec{p}$
the generalized momentum and $e>0$ the ion charge. The cylindrical
symmetry about the direction of the magnetic field implies conservation
of the axial projection $L_z$ of the angular momentum
$\vec{L}=\vec{r}\times \vec{p}$, whose quantized values will be noted
$\hbar \Lambda$.  It is convenient to choose the origin of the coordinates
on the neutral atom and to work in the symmetric Landau gauge, in which the
vector potential has the form $\vec{A}=\frac{1}{2}{\vec B}
\times \vec{r}$.
With this choice, after some simple algebra the Hamiltonian takes in
cylindrical coordinates $\{ \rho,\varphi,z  \}$ the following form
\be
H=\frac{p_z^2}{2m}+\frac{p_\rho^2}{2m}+\frac{L_z^2}{2m\rho^2}+\frac{m \omega^2}{2}  \rho^2 - \omega \L_z+V
\ee
of a particle in a diamagnetic potential of Larmor angular frequency
$\omega = \frac{eB}{2m}$ interacting with a fixed center of force $V$.
The coordinate $\rho$ is the distance from the $z$ axis, $p_\rho$ is its
conjugate momentum and $p_z$ is the linear momentum in the direction of
$\vec B$.

Conservation of the axial angular momentum reduces the orbital Zeeman
term $\omega \L_z$ to a constant but its influence on the spectrum
of $H$ is essential. In fact, quantized energies for ion motion in the
plane orthogonal to $\vec B$, the Landau levels \cite{BK-1999-LDL},
can be expressed in terms of $\Lambda$ and of the radial quantum number
$n$ by 
\be E_{n \Lambda}=\hbar \omega \left( 2n+|\Lambda| -
\Lambda +1  \right) \quad , \quad n=0,1,2, \cdots .  \label{landaulev}
\ee 
Eigenvalues have therefore infinite degeneracy, a quantum counterpart
to the classical freedom to choose the cyclotron orbit center keeping
the energy fixed. The level spacing  $2\hbar \omega$ for $n \to n\pm 1$
transitions corresponds to the energy of the classical cyclotron motion
with angular frequency twice the Larmor frequency of precession,
$\omega_c=eB/m=2 \omega$ .

The explicit expression of the eigenfunctions for planar motion with azimuthal
angle $\varphi$ is well known  (see, e.g., Ref.~\cite{BK-1999-LDL})
\be
\phi_{n \Lambda}\left( \rho , \varphi \right) = N_{n \Lambda}
y^{|\Lambda|} L_n^{|\Lambda|}\left( y^2 \right) e^{-y^2/2}  e^{i \Lambda
\varphi}  \quad ,  \quad y=\frac{\rho}{a_{\rm ho}},
\label{cylbasis}
\ee
where $a_{\rm ho}=\sqrt{\hbar/m\omega}$ is the harmonic oscillator length
and $N_{n \Lambda}$ a normalization constant. The quantum description
of helicoidal motion is finally completed by a uniform translation in
the $z$ direction
\be
\Phi_{n \Lambda}^E \left( \vec{r}  \right)=  \phi_{n \Lambda}\left( \rho , \varphi \right)
  |k_n |^{-1/2}
e^{i k_n z} .
\ee
Note that to simplify the notation we have defined $k_n=k_{n \Lambda}$
dropping the conserved quantum number $\Lambda$.  Conservation of the
total energy $E$ determines the magnitude of the longitudinal wave vector
\be
E=\frac{\hbar^2 k_n^2}{2m}+E_{n\Lambda}.
\label{energy}
\ee  
where the first term on the rhs is the kinetic energy in channel $n$, $E_{\rm kin}=\frac{\displaystyle \hbar^2
k_n^2}{\displaystyle 2m}$. 

Classically, as the atom and the ion approach, the spherically
symmetric interaction distorts the cyclotron orbit modifying the energy
distribution between longitudinal and transverse degrees of freedom.
Quantum mechanically, a Landau collisional state $\Phi_{n \Lambda}^E$
can be either reflected or transmitted across the collision center with
different probability amplitudes. In general, a collision starting with
initial quantum number $n$ will populate all states $\Phi_{n^\prime \Lambda}^E$
energetically accessible at energy $E$ with a transition probability
amplitude $f_{n^\prime n}$. These energetically open channels are characterized
by $k_{n^\prime}^2>0$, see Eq.~(\ref{energy}). The closed channels $k_{n^\prime}^2<0$
are not energetically allowed at large atom-ion separation but can be
populated during the collision giving rise to dramatic resonance effects.

Inspection of the typical length scales in the problem shows the essential
role played by Landau quantization for cold atom-ion collisions.
In the ultracold regime, the range of the interaction can be defined
in terms of the polarizability coefficient $C_4$ through $R^*=\sqrt{2 m
C_4 \hbar^{-2}}$ \cite{2009-ZI-PRA-010702}. The $R^*$ parameter is proportional
to the location $R^*/\pi$ of the last node in the zero-energy
wavefunction for infinite scattering length \cite{1996-ET-JRNIST-505}. As such,
it gives an estimate of the distance beyond which the wave function
oscillates freely. The characteristic energy $E^*=\hbar^2/(2 m R^{*2})$
sets the scale at which quantum effects become important \cite{1989-PSJ-JOSAB-2257}.
The quantum length scale associated to cyclotron motion is $a_{\rm ho}$
which corresponds to the size of the lowest energy Landau state. The
Landau levels spacing is finally given by $ \hbar \omega_c$.

Let us now consider a magnetic field intensity $B = 500$~G.
Tab.~\ref{table1} shows the characteristic quantities for some systems
of current or possible experimental interest at this value of magnetic
field. One can remark that the range of the interaction is comparable
to the harmonic oscillator length. Moreover, the transition between
the weak $R^* \ll a_{\rm ho}$ and the strong field $R^* \gg a_{\rm ho}$
limits can be explored by varying $B$ over an experimentally accessible
range. The Landau splitting of the order of $\mu$K makes the discretized
nature of the continuum in the magnetic field essential if collision
energies of this order of magnitude are considered.

\begin{table}[t]
\begin{center}
\caption{
Characteristic energy and length scales as defined in the text for few
sample atom-ion systems in a magnetic field $B = 500$~G. The oscillator
length at this value of the field is $a_{\rm ho}=3066 a_0$. Calculations
use the reduced mass of the atom-ion dimer.
} \label{table1} \vskip
12pt
\begin{tabular}{|l |  c c c |}
\hline \hline
{\rm System}  & $R^* (a_0)$ & $\hbar \omega_c(\mu{\rm K})$ & $E^*(\mu {\rm K})$ \\
\hline
{\rm Na+Ca}$^+$ & $2081$  & 2.5   &  1.4   \\
{\rm Rb+Ba}$^+$ & $4765$  & 0.70   &  0.92   \\
{\rm Yb+Yb}$^+$ & $5572$  & 0.43   &  1.1     \\

 \hline \hline
\end{tabular}
\end{center}
\end{table}

The scattering wavefunction presents the asymptotic $|z| \to \infty$ behaviour
\bea
\Psi_{n \Lambda}^{E}(\vec{r}) \to \Phi_{n \Lambda}^E(\vec{r}) + \sum_{k_{n^\prime}>0}
 f_{n^\prime n} \left( {\rm sign}(z)  k_{n^\prime} \leftarrow  k_n \right)
\Phi_{n^\prime  \Lambda}^E \left( \rho, \varphi , |z| \right)    
\label{asymptotic}
\eea
in terms of an incoming $\Phi_{n \Lambda}^E$ Landau state and of scattered waves outgoing in
the open channels. 
One should note that since the transverse motion is confined by the
magnetic field, scattering is effectively one dimensional. The measurable
reflection and transmission coefficients for a progressive $k_n>0$
incoming wave can be promptly identified from Eq.~(\ref{asymptotic}):
\be
R_{n^\prime n}= \left |  f_{n^\prime n} \left(   -k_{n^\prime} \leftarrow  k_n \right)  \right|^2 \quad  , \quad
T_{n^\prime n}= \left | \delta_{n^\prime n}  +  f_{n^\prime n} \left(   k_{n^\prime} \leftarrow  k_n \right)  \right|^2 .
\ee

A different choice of the standardization of the wavefunction for
$|z| \to \infty$, the standing wave boundary conditions, will be more
convenient for numerical calculations. In order to make explicit use of
the invariance of the problem under the coordinate inversion $\vec{r}
\to -\vec{r}$ we first define new sets of reference functions in the $z$
motion with well defined behavior under the $z\to -z$ reflection
$\pi_z$. Namely, diagonal matrices of functions ${\mathbf F}^{\pi_z}$
and ${\mathbf G}^{\pi_z}$ regular and irregular at the origin are defined for open channels and parity ${\pi_z}=\pm$
by 
\bea
\left \{
\begin{tabular}{l}
$F_n^+(z)= k_n^{-1/2}  \cos(k_n z)$  \\
$F_n^{-}(z)= k_n^{-1/2}  \sin(k_n z)$ 
\end{tabular}
\right.
\quad , \quad
\left \{
\begin{tabular}{l}
$G_n^+(z)= k_n^{-1/2}  \sin(k_n |z|)$ \\ 
$G_n^{-}(z)= -k_n^{-1/2} {\rm sign}(z) \cos(k_n z).$ 
\end{tabular}
\right.
\label{asy1}
\eea
The corresponding reference functions for closed channels are
\bea
\left \{
\begin{tabular}{l}
$ F_n^+(z)=   \cosh(k_n z)  $  \\
$ F_n^{-}(z)=   \sinh(k_n z)  $
\end{tabular}
\right.
\quad , \quad
\left \{
\begin{tabular}{l}
$ G_n^+(z)=   -\exp(-k_n |z|) $ \\
$ G_n^{-}(z)=   -{\rm sign}(z) \exp(-k_n |z|).  $
\end{tabular}
\right.
\label{asy4}
\eea

With these definitions, total parity of product functions $\phi_{n
\Lambda}F_n^\pm$ and  $\phi_{n \Lambda}G_n^\pm$ is $p=\pi_z (-1)^\Lambda$.
We can now define a real reactance matrix ${\mathbf K}^{\pi_z}$ for each
$\pi_z=\pm$ and angular momentum $\Lambda$ according to the asymptotic
behaviour~\cite{BK-1964-MLG}
\be
\Psi_{n \Lambda}^{Ep}(\vec{r}) \to \phi_{n \Lambda}\left( \rho, \varphi \right) F_n^\pm(z) - \sum_{n^\prime}
 K_{{n^\prime} n}^{\pi_z}
\phi_{{n^\prime}  \Lambda}\left( \rho, \varphi \right)   G_n^\pm (z) \quad , \quad
|z| \to \infty.
\label{kmat}
\ee
If the transition matrix is also decomposed into symmetric and antisymmetric
components ${\mathbf{f}}^{\pi_z}$,
\be f_{n^\prime n} \left(
{\rm sign}(z)  k_{n^\prime} \leftarrow  k_n \right)= f_{n^\prime n}^+(E)
+ {\rm sign}(z) f_{n^\prime n}^-(E)
\ee
the matrix relations
\be 
\mathbf{f}^{\pi_z} = i \mathbf{K}^{\pi_z} \left( \mathbf{I} - i \mathbf{K}^{\pi_z}  \right)^{-1}  
\ee
obtained comparing (\ref{asymptotic}) and (\ref{kmat}) can then be used
to determine the $\mathbf{f}^{\pi_z}$ in terms of ${\bf K}^{\pi_z}$.
Finally, we will make use below of the unitary and symmetric scattering matrix,
defined for the two parities as 
$\mathbf{S}^{\pi_z} = \left( \mathbf{I} + i \mathbf{K}^{\pi_z}  \right)  \left( \mathbf{I} - i \mathbf{K}^{\pi_z}  \right)^{-1}$.

\subsection{Wigner laws}

The scattering quantities obey at low energy Wigner laws, valid
when the asymptotic de Broglie wavelength is much larger than
the range of the effective potential in a given channel. Following
Ref.~\cite{1960-LMD-NUP-275}, a collision channel in which the collision
energy tends to zero will be termed {\it new}. Similarly to the 3D
case, in which Wigner laws depend on the angular momentum quantum
number, even in the present 1D configuration it is convenient to treat
separately the ${\pi_z}=\pm $ parity cases. An analysis along the lines
of Ref.~\cite{1960-LMD-NUP-275} gives the asymptotic behaviour,
\bea
f_{nn}^+ \sim -1 + O(k_n) \quad , \quad 
f_{nn}^- \sim  k_n .
\eea   
implying that for vanishing collision energy at the $n$-th collision
threshold a Landau state will be reflected with unit probability,
irrespective of the value of $\Lambda$. Moreover, the odd transition
matrix element becomes negligible with respect to the even component ,
such that particles will be scattered with equal probability $|f^+_{nn}|^2$
in forward and backward directions.

Inelastic processes consist in a transition to a Landau state of different
transverse energy, and the corresponding scattering amplitudes obey 
for both parities the asymptotic law
\bea
f_{n^\prime n}^ \pm \sim \sqrt{k_n} ,
\eea   
if $n$ is new. The transition probability to a newly energetically
accessible channel is then vanishing at threshold  both in reflection
and transmission.
A similar relation 
\bea
f_{n^\prime n}^ \pm \sim \sqrt{k_{n^\prime}} 
\eea 
holds in the case where $n^\prime$ is new.

The Wigner laws allows one to define for positive parity an effective coupling
constant $g_{\rm 1D}(k_n)$ finite at threshold ($k_n \to 0$), which
embeds the details of the short range dynamics in a single parameter
\be
f_{nn}^+=- \frac{\displaystyle 1}{ \displaystyle 1 - i \frac{\displaystyle \hbar^2 k_n}{\displaystyle 2 m g_{\rm 1D}}}
\ee
Such $g_{\rm 1D}(k_n)$ can be used for instance to parametrize a 1D
pseudopotential $V_p$, {\it i.e.} a zero-range interaction $V_p=g_{\rm
1D} \delta(z)$ which generates the same elastic scattering amplitude
as the full problem at momentum $k_n$. An associated one dimensional
scattering length can be defined for one open channel as the limit
value $a_{\rm 1D} = -\lim_{k\to 0} \eta/k$, where $\eta(k)$ is the
scattering phase shift extracted from the asymptotic wave function
$\Psi_{00}^{E+}(\vec r) \propto \sin (k |z| + \eta) \phi_{00}\left(
\rho \right)$~\cite{1998-MO-PRL-938}.  Its relation with the scattering
amplitude is
\be
\frac{a_{\rm 1D}}{a_{\rm ho}}=-\frac{\hbar \omega a_{\rm ho}}{g_{\rm 1D}}.
\label{gefa}
\ee
For a purely elastic collision $g_{\rm 1D}$ is real, otherwise its
non-vanishing imaginary part accounts for inelastic loss processes.
Note that unlike the three-dimensional case, where $a_{\rm 3D}$ is
proportional to $g_{\rm 3D}$, here the scattering length is inversely
proportional to the coupling constant.  We will restrict most of
our discussions to $g_{\rm 1D}$, since this is the quantity usually measured in
experiments \cite{2004-BP-NAT-277}. Moreover, pseudopotentials with coupling strength $g_{\rm
1D}$ are often used to model two-body interactions in many-body theories.
A numerically exact approach for the determination of the scattering
amplitude and hence of the coupling constant is the subject of section
\ref{numm}.

\subsection{Feshbach resonances in quasi 1D}

Feshbach resonance theory is well established for three dimensional
collisions. We shortly recall here its main elements in order
to point out some differences with respect to the present
quasi one-dimensional configuration. Let us focus for definiteness
on the case of one open channel and one isolated resonance. We will consider
positive parity scattering and drop the $\pi_z$ parity label for notational convenience. A formal
separation of background and resonant contributions to the scattering
matrix can be performed following the standard Fano-Feshbach
approach~\cite{1958-HF-ANNP-357,1961-UF-PRA-1866}
\be
S=S_{\rm bg} \frac{2(E-E_0-\zeta_E)-i\gamma_E}{2(E-E_0-\zeta_E)+ i \gamma_E} ,
\label{Feshbach}
\ee
where $S_{\rm bg}$ is the off-resonance scattering matrix, $E_0$ the
energy location of the molecular state responsible for the resonance,
$\zeta$ and $\gamma$ the resonance shift and width, respectively.
Near resonance the quantity $E_0$ typically presents a smooth linear
dependence on the external parameters of the problem, the magnetic field
or the potential depth in the present case.

At low momentum, according to the Wigner laws $S \sim (-1 -2 i k a_{\rm
1D})$, $S_{\rm bg} \sim (-1 - 2 i k a_{\rm 1D ,  bg})$, $\zeta_E \sim {\rm
const} $ and $\gamma_E \sim k$. Taking the $E\to 0$ limit, we obtain
the resonantly modified $ a_{\rm 1D}$ as
\be
 a_{\rm 1D} =  a_{\rm 1D , bg} \left( 1- \frac{\Delta}{E_0+\zeta_0}
 \right),
\ee 
where we have defined $\Delta = \lim_{k \to 0} \gamma_E / (2 k a_{\rm 1D ,
bg})$. Note that on resonance ($E_0 = - \zeta_0$) the 1D scattering length
diverges and according to (\ref{gefa}) the system becomes non interacting,
$g_{\rm 1D} \to 0 $ and the transmission coefficient tends to one. This
behaviour should be contrasted with the infinitely strong interaction
obtained in three dimensions as the molecular state crosses the collision
threshold.  Note however that near resonance $a_{\rm 1D}$ will cross zero,
and $g_{\rm 1D}$ will diverge, if $E_0+\zeta_0=\Delta$. With this remark,
in the following we will refer to the divergence of the effective coupling
constant as to a zero-energy resonance situation.

\section{Numerical methods}
\label{numm}

We solve our problem in the framework of the time independent close
coupling approach to potential scattering. In this method the full
wavefunction is expanded in a suitable complete basis depending on a
reduced set of coordinates. The choice of the basis function is inspired
by the physics of the problem.  Let us define a spherical system of
coordinates $\{ r,\theta,\varphi \}$ with origin on the collision center
and polar axis in the $z$ direction. We identify an ``inner region''
$r \leq R_0$ where the isotropic interatomic interaction initially
dominates. The criterion for choosing the boundary $R_0$ is given
below. In this internal region an expansion in spherical harmonics is used
\be
\Psi^{Ep}_{n \Lambda}({\vec r}) = {\sum_{\ell \geq |\Lambda|}}^\prime
r^{-1} f_{\ell n}(r) Y_{\ell \Lambda}({\hat r}),
\label{spherical}
\ee
where the sum $\sum^\prime$ runs only over even or odd values of $\ell$
for positive or negative parity, respectively.
The coupled channel Schr{\"o}dinger equation takes the form 
\be
\left[ -{\mathbf I}  \frac{\hbar^2}{2m} \frac{d^2}{dr^2} +{\mathbf
v}+{\mathbf C} - \Lambda \omega {\mathbf I}   \right] {\mathbf f}(r)=E{\mathbf f}(r) .
\label{ccsph}
\ee
The effective centrifugal potential $\mathbf v$ has a simple diagonal form with elements
\be
v_\ell(r)=\frac{\hbar^2 \ell (\ell+1)}{2mr^2}+V(r)
\ee
whereas the coupling matrix $\mathbf C$ is formed by the matrix elements of the transverse
oscillator potential
\be
C_{\ell^\prime \ell}=\frac{m \omega^2 r^2}{2} \langle Y_{\ell^\prime \Lambda} | \sin^2(\theta) | Y_{\ell \Lambda}  \rangle.
\ee
Eq.~(\ref{ccsph}) is solved up to $R_0$ using the Johnson-Manolopoulos
propagator \cite{1986-DEM-JCP-6425}, which computes the log derivative matrix ${\mathbf
Z}=\left( \frac{d}{dr}{\mathbf f} \right) {\mathbf f}^{-1}$ along a reaction coordinate using embedding
propagators.

In the external $r>R_0$ region it is more natural to use
cylindrical coordinates and an orthogonal basis of Landau oscillator states, 
Eq.~(\ref{cylbasis})
\be
\Psi^{Ep}_{n \Lambda}({\vec r}) = \sum_{n^\prime}
 h_{n^\prime n}(z) \phi_{ n^\prime \Lambda}(\rho,\varphi).
\label{cylindrical}
\ee
The coupled channel Schr{\"o}dinger equation in the cylindrical basis becomes
\be
\left[ -{\mathbf I} \frac{\hbar^2}{2m  } \frac{d^2}{dz^2 }  +{\mathbf
E_L}+{\mathbf u}   \right] {\mathbf h}(z)=E{\mathbf h}(z).
\label{cccyl}
\ee
The diagonal matrix $\mathbf E_L$ is composed by the Landau levels
(\ref{landaulev}) and coupling $\mathbf u$ arises from the spherical
atom-ion interaction $V$.  In practice we will always switch to the
cylindrical representation at distances where the $V(r)=-C_4r^{-4}$
asymptotic form is accurate. Moreover, in the same region the size of
transverse oscillator states is much smaller than the radial coordinate $r$, such
that a series expansion in the small parameter $a_{\rm ho}/r$ can be performed.
We find matrix elements
\be
u_{n^\prime n}=\langle \phi_{n^\prime \Lambda} | V  |  \phi_{n
\Lambda}  \rangle = - \frac{C_4}{z^4} \left( 1  - \frac{2}{z^2} \langle
\phi_{n^\prime \Lambda} | \rho^2  |  \phi_{n \Lambda} \rangle + \cdots
\right).
\ee
The second term in the expansion is rapidly decreasing $\sim z^{-6}$
and in the present calculations we only retain lower order $\sim z^{-4}$ diagonal coupling
elements $u_{nn}$ for both open and closed channels.

The representations (\ref{spherical}) and (\ref{cylindrical}) are matched
on the spherical surface $S_0$ defined by $r=R_0$. We find that to
insure optimal numerical convergence one should choose $R_0$ in such a
way that the relevant Landau states are localized in the two hemispheres
without significant overlap on the equatorial plane \footnote{Typically,
$R_0=10a_{\rm ho}$ insures a good convergence for collisions starting
in the lowest Landau levels.}.  The matching procedure is analogous
to the one used in reactive scattering \cite{1989-JML-CPL-178,
1987-RTP-JCP-3888}.  We anticipate that the change from spherical to
cylindrical geometry is at the origin of the peculiar scattering and
resonance phenomena described in the next section.

In the external region $r>R_0$  we integrate inward the equations
(\ref{cccyl}) from very large values of $z$, with starting value
the asymptotic boundary condition (\ref{asy1})-(\ref{asy4}). Let us
denote the corresponding matrix solutions $\hat{\mathbf F}^\pm$ and
$\hat{\mathbf G}^\pm$, which retain diagonal form due to the neglect of
the off diagonal elements in the coupling $\mathbf u$. The scattering
solution is represented as
\be
\Psi^{Ep}_{n \Lambda}({\vec r}) = \sum_{n^\prime}
 \left[ \hat{F}^\pm_{n^\prime}(z) \delta_{n^\prime n} - \hat{G}^\pm_{n^\prime }(z)   K^{\pi_z}_{n^\prime n}   \right] \phi_{n^\prime  \Lambda}(\rho,\varphi) \quad , \quad r>R_0.
\ee

Continuity of the scattering wavefunction and of its normal derivative
on the spherical surface $S_0$ is used to determine $\mathbf K$ in terms of the
logarithmic derivative matrix $\mathbf Z$. The projection matrices
\bea
  {\cal F}_{\ell n}= \langle  Y_{\ell \Lambda}   | {\hat F}_n \phi_{n
  \Lambda}  \rangle_{S_0}  \nonumber  \\ 
   {\cal F}_{\ell n}^\prime
  =  \langle  Y_{\ell \Lambda}   | \frac{\partial}{\partial r}
  ({\hat F}_n \phi_{n \Lambda})  \rangle_{S_0}
\label{proj}
\eea
needed to this aim \cite{1987-RTP-JCP-3888} are numerically evaluated
using Gauss-Legendre quadratures. Similar projection matrices
${\cal G}_{\ell n}$ and $ {\cal G}_{\ell n}^\prime$ involving the irregular
reference functions are defined by replacing $\hat F$ with $\hat G$
on the rhs of (\ref{proj}).
The reactance matrix is finally obtained for each parity using the
equation \cite{1987-RTP-JCP-3888}
\be
{\mathbf K}=\left( \mathbf{Z}    \mathbf{\cal G}  -  \mathbf{\cal G}^\prime  \right)^{-1} 
         \left(   \mathbf{Z}   \mathbf{\cal F} -\mathbf{\cal F}^\prime   \right).
\ee

\section{Results and discussion}

In this section we apply our formalism to atom-ion diffusion with $\Lambda
=0$, a somewhat idealized situation to test our theoretical formalism
which can however be in principle experimentally realized. For instance,
instantaneous switching of a magnetic field in the propagation direction
of an ion beam would project the plane wave describing the beam on
a superposition of Landau states with different $n$ and $\Lambda =0$
without affecting the longitudinal momentum $k$. A successive measurement
of the transverse energy could be used to select the desired initial
asymptotic state $n$ for the collision.

The default molecular parameters we choose for our model correspond to the Na+Ca$^+$
system. However, previous work has shown that the short range molecular
physics can be accurately described in the ultracold regime by a single
quantum defect parameter independent of energy and angular momentum,
which can be identified with the $s$-wave three-dimensional scattering
length $a_s$ \cite{2009-ZI-PRA-010702}.  Within this approximation,
our results will be cast in an essentially  system independent form by
presenting them in terms of the only independent dimensionless groups
\be
 \frac{a_s}{R^*} \quad ,\quad  \frac{E}{\hbar \omega_c} \quad  ,\quad  \frac{a_{\rm ho}}{R^*} ,
\ee
that can be built using the physical parameters of the problem. Note
that in this work we set $m$ equal to the reduced mass of the system, for
future comparison with the free-space results ~\cite{2009-ZI-PRA-010702}.

We first consider collisions starting in the lowest Landau level $n=0$
at zero collision energy and calculate the effective coupling constant
$g_{\rm 1D}(k_0 \to 0)$.  We make the three dimensional scattering
length $a_s$ vary from infinite negative to positive infinite values
by artificially modifying the interaction potential. The variation of
$a_s$ in units of $a_{\rm ho}$ will be mapped to a finite interval by
introducing the normalized quantum defect $\xi=(2/\pi) \arctan(a_s/a_{\rm
ho})$. This analysis will be repeated for three values of $a_{\rm ho}$.
We will then analyze the behaviour with the collision energy of the
scattering observables for some selected values of $a_{\rm ho}$ and $a_s$.

\subsection{Weak field regime}

\begin{figure}[h!]
\includegraphics[width=\columnwidth,clip]{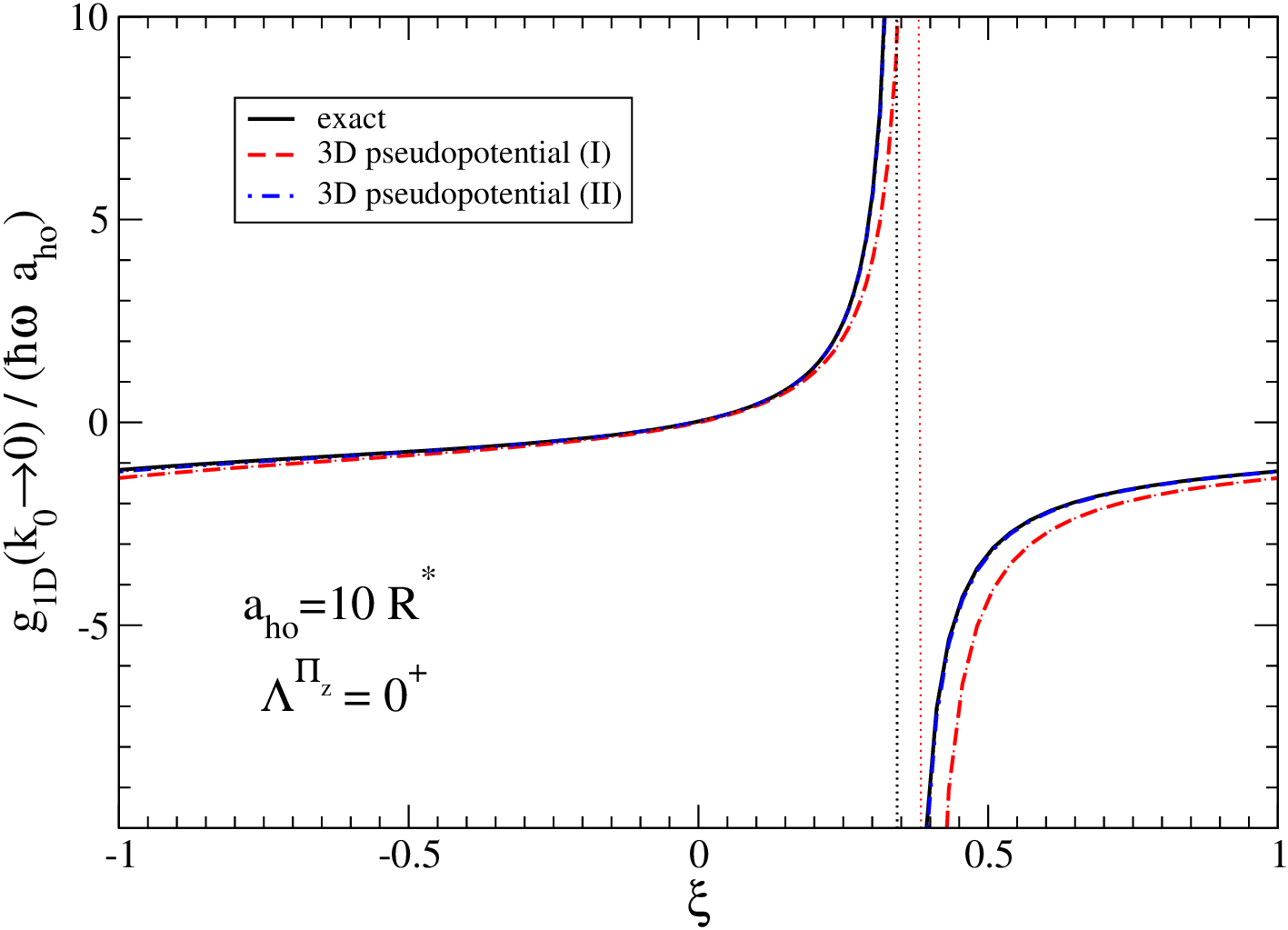}
\caption{Numerically computed effective coupling constant for Ca$^+$+Na collisions in a
weak magnetic field with $a_{\rm ho}=10R^*$ as a function of the normalized
scattering phase $\xi=(2/\pi) \arctan (a_s/a_{\rm ho})$ for axial
angular momentum $\Lambda^{\pi_z}=0^+$ (full line). Vertical dotted lines denote the position
where $g_{\rm 1D}$ diverges ($a_{\rm 1D}=0$).
Energy independent end energy dependent pseudopotential approximations (see text) 
labeled I and II respectively are also shown.
}
\label{fig1}
\end{figure}

Let us first fix a value of $B$ such that the size of the lowest Landau state
is significantly larger than the interaction range. We will refer to
this condition as the weak field limit, and choose more specifically
$a_{\rm ho} =10   R^*$ ($B=10.9$~G). A salient feature observed in Fig.~\ref{fig1}
is the strong divergence of the effective coupling constant at a
characteristic value of $a_s$. That is, the effective 1D coupling
strength can be tuned to virtually any value by modifying the value of
$a_s$ with respect to $a_{\rm ho}$, for instance through a variation of the applied
magnetic field. We stress that this zero energy resonance arises from
the quantization of the continuum in Landau levels and does not depend on
resonance phenomena preexisting in free space. As such it can be termed
a geometric resonance effect.

Interestingly, since in the weak field limit the range of the
potential can be neglected with respect to the transverse extension
of the Landau state,² the interaction $V$ can be expected to be well
described by a zero range 3D pseudopotential. With this approximation,
Eq.~(\ref{cccyl}) for $\Lambda=0$ reduces to an analytically solvable
model~\cite{1998-MO-PRL-938,1995-TPG-PRA-607}.
The effective coupling constant of Ref.~\cite{1998-MO-PRL-938} is
\be
\frac{g_{\rm 1D}}{\hbar \omega a_{\rm ho}}=2\frac{a_s}{a_{\rm ho}} \left( 1+ \frac{a_s}{a_{\rm ho}}  \zeta(1/2)  \right)^{-1} ,
\label{olshanii}
\ee
where $\zeta$ is the Riemann zeta function and $\zeta(1/2)=-1.46035$.
The zero-order result $g_{\rm 1D}=2 a_s \hbar \omega$, obtained
by neglecting the second term in the parenthesis on the rhs,
simply represents the renormalization of $a_s$ obtained in the Born
approximation by averaging the three-dimensional pseudopotential over
the initial transverse Landau state.  However, Eq.~(\ref{olshanii})
also shows that a confinement-induced resonance arising from virtual
excitations to asymptotically closed Landau states should appear for $
a_s=-a_{\rm ho}/ \zeta(1/2) \simeq 0.6848 ~ a_{\rm ho}$.

This nontrivial prediction is confirmed by the comparison with our
numerical calculation shown in Fig.~\ref{fig1}. The small shift in
the resonance position can be understood by realizing that the result
(\ref{olshanii}) is exact only in the limit of vanishingly small {\it
axial} collision energy. However, when the atom and the ion are close
together the relative energy $E$ in their unconfined 3D motion does
not vanish but is rather equal to the zero point energy $\hbar \omega$,
see Eq.~(\ref{landaulev}). This correction can be introduced by using
an energy dependent 3D peudopotential \cite{2002-ELB-PRA-013403},
which replaces the zero energy parameter $a_s$ with its finite
energy counterpart $-\tan(\delta_{\rm 3D})/k_{\rm 3D}$. Here
$\delta_{\rm 3D}$ is the $s$-wave phase shift at energy $E$ and $k_{\rm
3D}=\sqrt{2mE/\hbar^2}$ is the wave vector in the three-dimensional
motion. With this modification our numerical calculation becomes virtually
identical to Eq.~(\ref{olshanii}) thus confirming the accuracy of the
pseudopotential approximation even for a long-range $r^{-4}$ potential
under the appropriate conditions.
\begin{figure}[hb!]
\includegraphics[width=\columnwidth,clip]{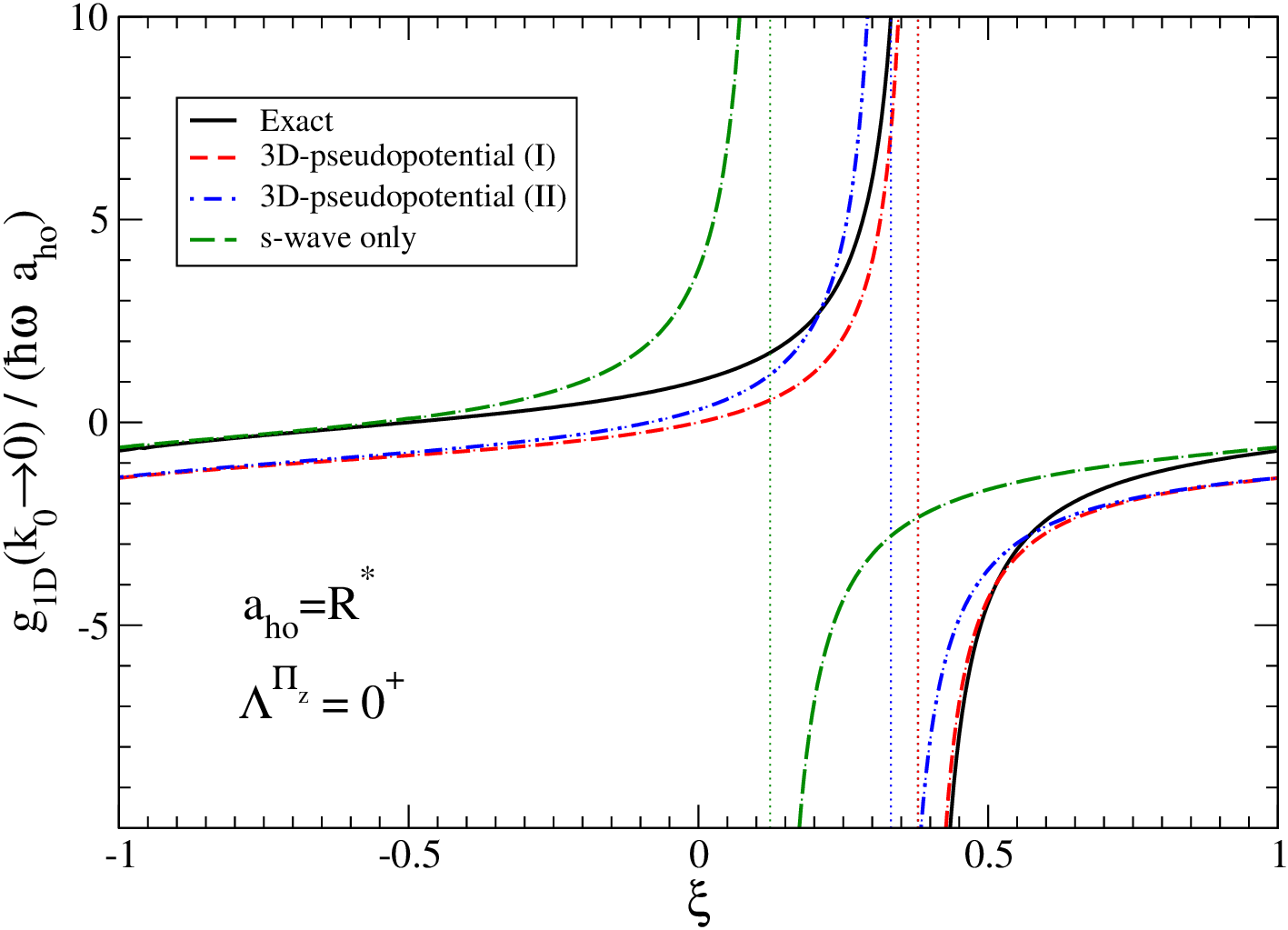}
\caption{Same as figure \ref{fig1} but in the intermediate field case $a_{\rm ho}=R^*$.
The result of a purely $s$-wave calculation with the full molecular potential is also shown (see text)
.}
\label{fig2}
\end{figure}

Since the zero point energy is small with respect to the centrifugal
barriers in Eq.~(\ref{ccsph}) one can expect, as indeed proven by the
agreement with the $s$ wave model (\ref{olshanii}), that diffusion in
partial waves $\ell> 0$ should be significantly suppressed.
Our numerical approach allows us to check this property by a direct
calculation. To this aim, we set $V=0$ in the $v_{\ell>0} $ components
of the interaction matrix $\mathbf{v}$. The non $s$-wave potentials
are also turned off in the external region by using in the projection
matrix elements (\ref{proj}) with $\ell>0$ the unperturbed waves $F$ and $G$
in the place of $\hat F$ and $\hat G$.
Note that in spite of the fact that dynamically a single partial wave
contributes to scattering, one cannot simply use a unique $\ell=0$
in the expansion (\ref{spherical}), since a sufficient number of partial waves
is in any instance cinematically necessary in order to represent the
relevant Landau states on the spherical surface $S_0$ of radius $R_0$.  The result of
this purely $s$-wave calculation is not included in Fig.~\ref{fig1}
as it is virtually indistinguishable from the exact result.

\subsection{Intermediate field regime}

Let us now consider the variation of $g_{\rm 1D}$ with $a_s$ in the
intermediate field regime $a_{\rm ho} = R^*$ ($B=1.09 \times 10^3$~G).
Like in weak field conditions, a single geometric resonance is
observed in the numerical simulation, see Fig.~\ref{fig2}. The pseudopotential result in
Eq.~(\ref{olshanii}) in the energy-independent (I) and energy-dependent (II)
formulations is only qualitatively correct.  The purely $s$-wave calculation
shows a significant shift in the resonance position from both the exact
calculation and the pseudopotential models.

These findings can be rationalized as follows. The discrepancy between
the pseudopotential and the purely $s$-wave result demonstrates the
breakdown of the assumption of a zero-range potential, that is the
importance of the long-range interplay between atom-ion and magnetic field
potentials. The disagreement between the $s$-wave and the exact result
is due to the important contribution of higher order partial waves. This
contribution does not vanish even in the $k\to 0$ limit because of the
zero point energy. Its value ($\simeq 3 \mu$K), a factor $10^2$ larger
than in the weak field case, and the long range nature of the potential
insure that a few partial waves are scattered from the potential. More
quantitatively, by artificially varying the number of partial waves
as described above we find that a calculation restricted to $\ell=0-4$
determines with high accuracy  $g_{\rm 1D}$. The good agreement in the
resonance position between the exact result and the pseudopotential (I)
approximation should be regarded as merely coincidential, stemming from
the compensation of long-range and higher order partial waves effects.

\begin{figure}[h!]
\includegraphics[width=\columnwidth,clip]{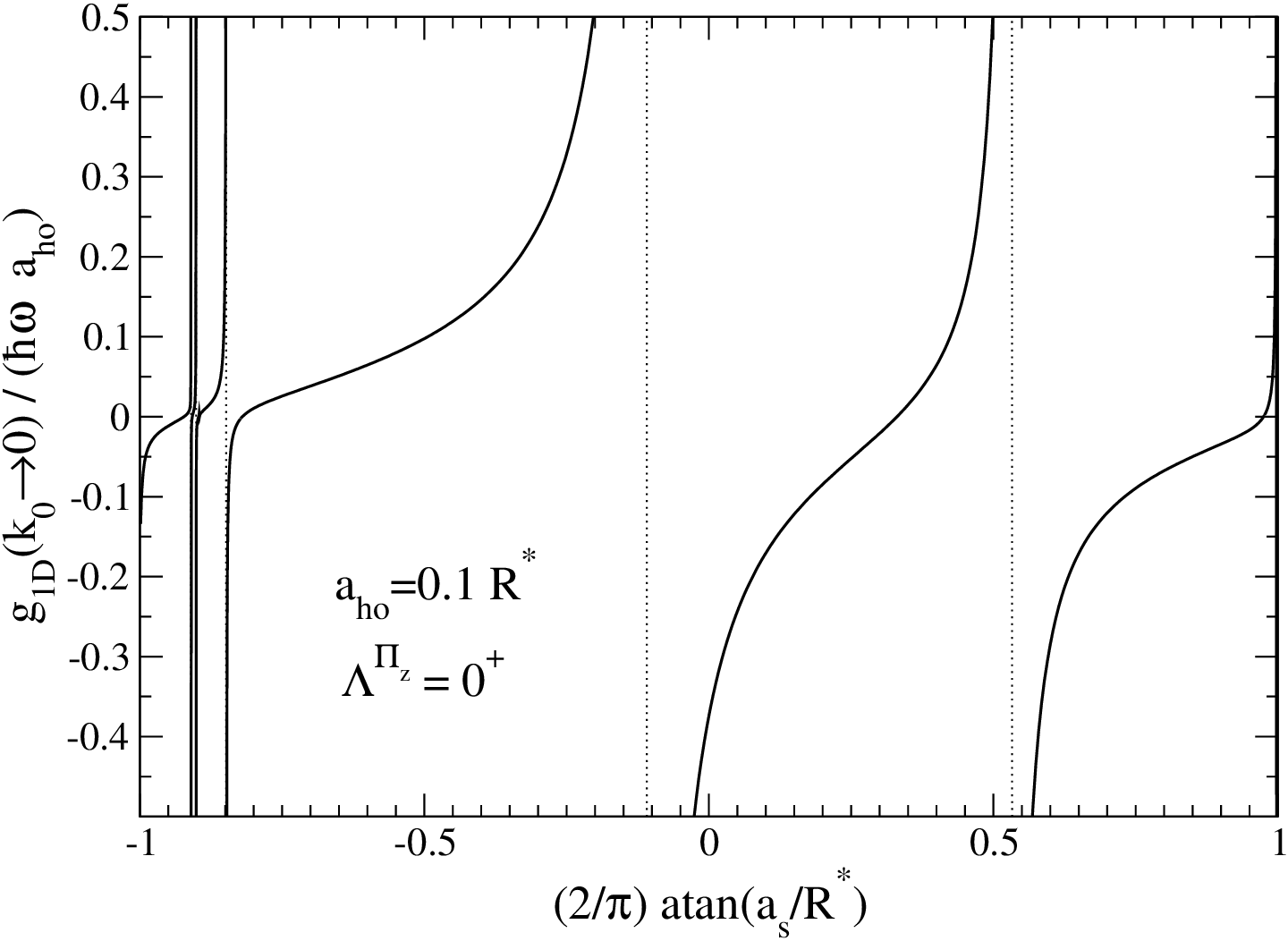}
\caption{Same as figure \ref{fig1} but in the strong field case $a_{\rm ho}=0.1 R^*$.
The pseudopotential approximation breaks down in this situation and is not shown. 
Note that for graphical purposes the normalized quantum defect on the horizontal axis is defined in terms
of $R^*$ rather than of $a_{\rm ho}$.
}
\label{fig3}
\end{figure}
\subsection{Strong field regime}
Finally, Fig.~\ref{fig3} shows the behaviour of  $g_{\rm 1D}$ for a
strong field case $a_{\rm ho} = 0.1 R^*$, a tight confinement condition
in which the potential range is larger than the Landau orbit radius. The
required magnetic field intensity is of $B=1.09\times  10^5$~G, a
large yet experimentally attainable value. In this conditions both
the pseudopotential and the purely $s$-wave model completely fail to
reproduce the numerical results. Several zero-energy resonances are
now available as $a_s$ (or $B$) is made to vary. The nature of such
resonances can be understood from test calculations in which the atom-ion
potential is suppressed in selected partial waves. Such calculations
show that each broad feature in the $g_{\rm 1D}$ of Fig.~\ref{fig3}
arises from the combined effect of a relatively large number of partial
waves $\ell=0-12$. That is, due to the relatively large zero point
energy ($\hbar \omega \simeq 300 \mu$K) the wave function picks up
a significant phase shift at short range from different partial wave
effective potentials. The narrowest features observed near $\xi=-1$
can in contrast be assigned to a single or few $\ell \geq 4$ angular
momentum values, suggesting that most of the amplitude of the resonant
bound state is localized in the potential well region. An analysis of
the bound states of the systems will be needed to assign more precisely
the features observed in the present scattering calculations.
\begin{figure}[h!]  \includegraphics[width=\columnwidth,clip]{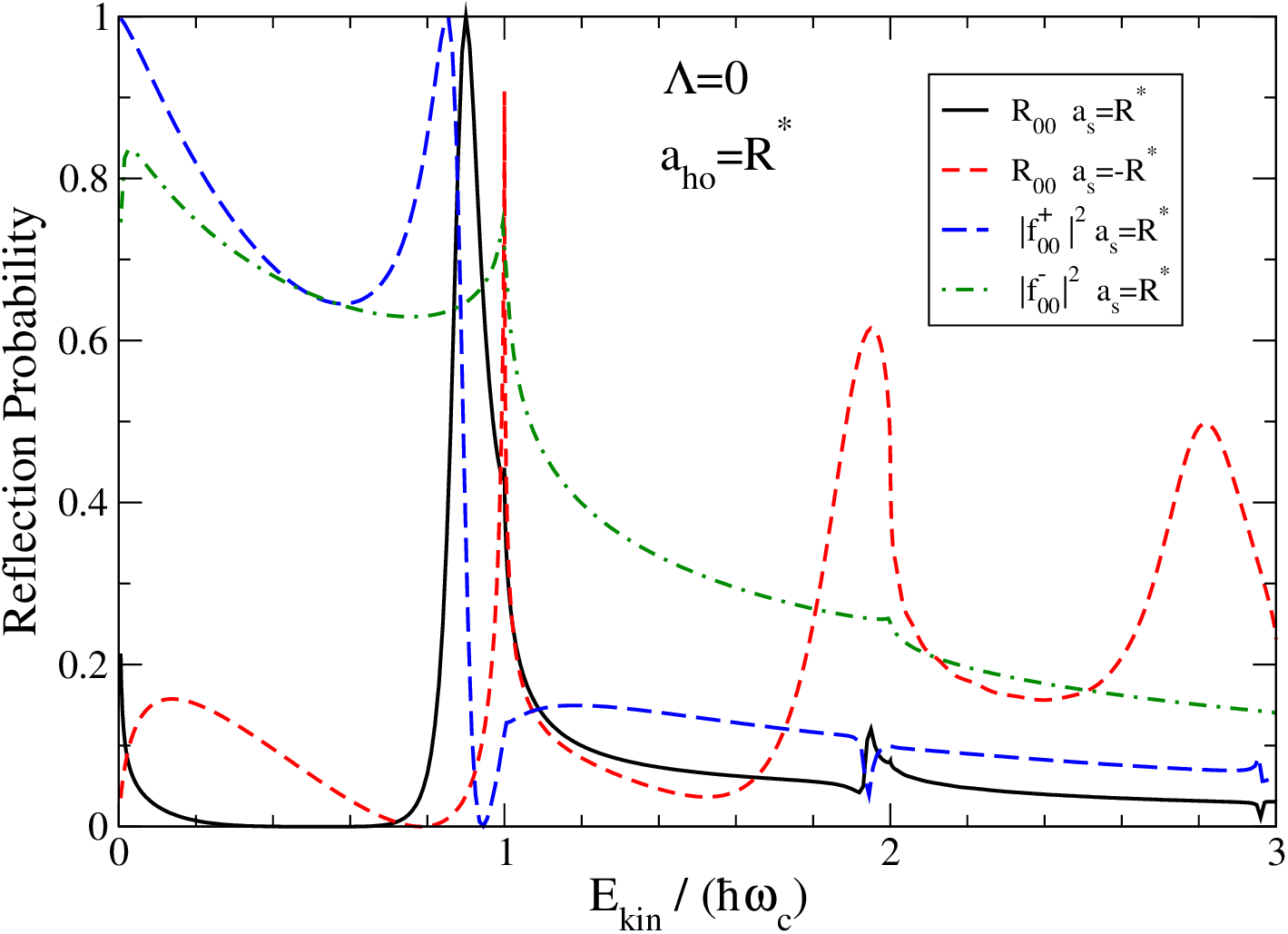}
\caption{Elastic reflection probability for an ion in the lowest
Landau level as a function of collision energy in the intermediate
field case $a_{\rm ho}=R^*$. Axial angular momentum is $\Lambda=0$ and
typical values of the tridimensional scattering length $a_s=\pm R^*$
are selected. Odd and even contributions to the reflection probability
are separately shown for the $a_s=R^*$ case. Resonance and threshold
phenomena in the reflection probability are discussed in the text.  }
\label{fig4} \end{figure}

\subsection{Behaviour with collision energy}

\begin{figure}[htbp]
\includegraphics[width=\columnwidth,clip]{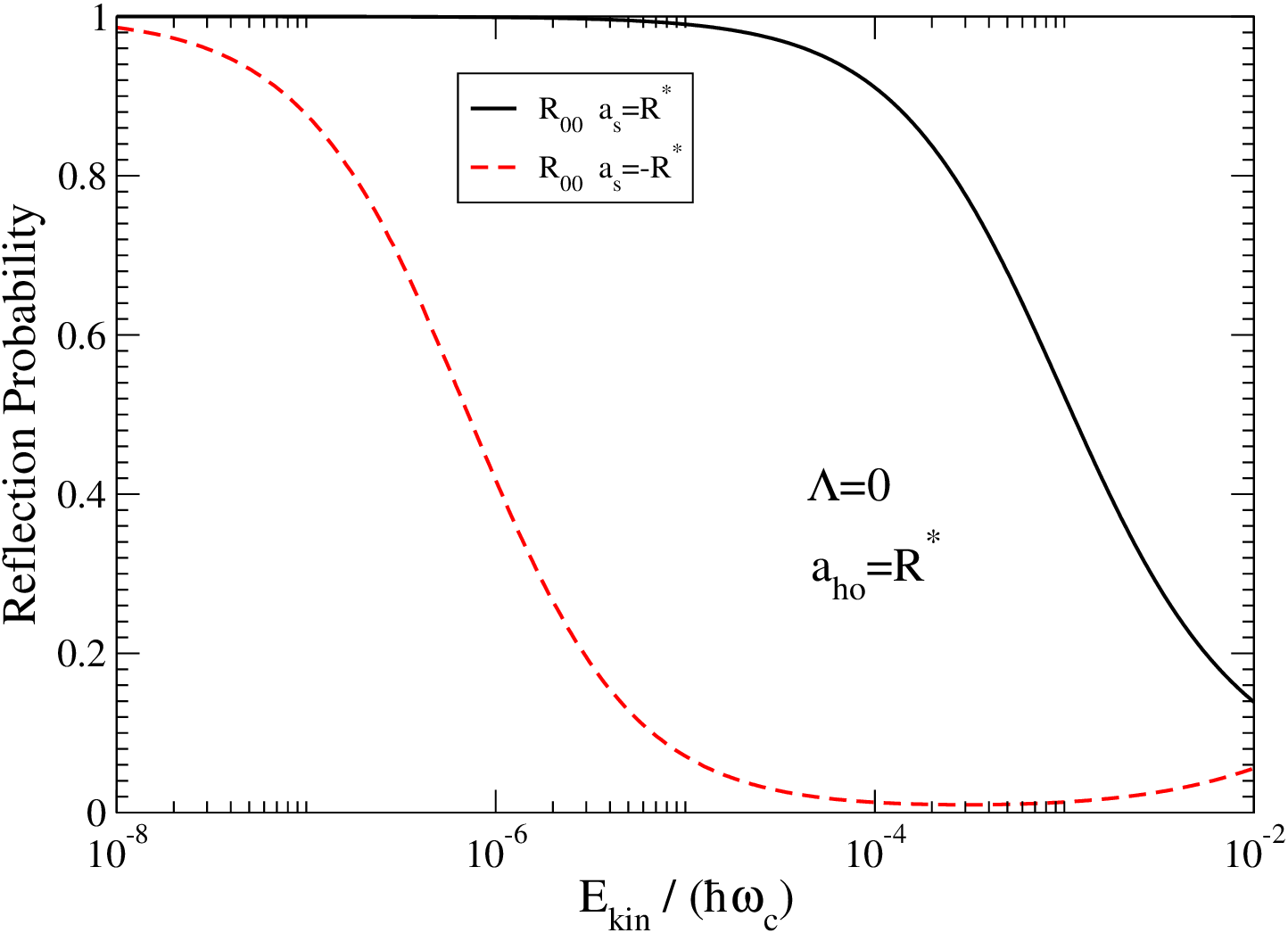}
\caption{
Near threshold elastic reflection probability for $a_s=\pm R^*$. Systems parameters are as in Fig.~\ref{fig4}. 
}
\label{fig5}
\end{figure}

The zero energy resonances described so far may shift as a function of
the potential strength $a_s$ into the scattering continuum giving rise
to possible resonance effects observable as a function of collision
energy. Let us consider in detail two potentials supporting ``typical''
scattering lengths $a_s=\pm R^*$ in the intermediate field case
$a_{\rm ho} =  R^*$. Fig.~\ref{fig4} shows the {\it elastic} reflection
probability $R_{00}$ for a scattering event starting with the ion in
the lowest Landau state $\phi_{00}$, as a function of collision energy.
A decomposition of the scattering amplitude in even and odd components
shows that the small value of the reflection coefficient
$R_{00}=|f^+_{00}  -f^-_{00} 
 |^2 $ between the $n=0$ and $1$ thresholds
is due to interference of parity amplitudes of relatively large magnitude,
see Fig.~\ref{fig4}. A similar situation has been reported for neutral
atom scattering from short-range potentials in \cite{2006-JIK-PRL-193203}.
Fig.~\ref{fig5} stresses the fact that the Wigner regime near the $n=0$
threshold at which $R_{00} \to 1$ is correctly reproduced by the numerical
calculation but is extremely narrow, for $a_s=- R^*$ in particular.

For $a_s=R^*$ a first peak at which $R_{00}$ reaches one is observed
below the opening of the $n=1$ level with energy $\hbar \omega_c$.
This feature arises from the positive parity part ${\mathbf f}^+$ of
the scattering amplitude, and can be supposed to be a resonance arising
from trapping of the ion in a long-lived superposition of excited
Landau states.

To confirm the resonant character of this feature we calculate the time
delay matrix introduced by Smith~~\cite{1960-FTS-PR-349}, defined for each parity as
\be
{\mathbf Q^{\pi_z}}=i \hbar \left( {\mathbf S^{\pi_z}}\right)^\dagger \frac{d {\mathbf S^{\pi_z}}}{dE} .
\ee
Note that $\mathbf Q$ is expressed in terms of the open channel
scattering matrix, such that its dimension equals the number of
energetically accessible channels. The evolution with energy of the
largest eigenvalue $q$ of the lifetime matrix can provide in particular
information on the lifetime of the complex formed during a collision. For
instance, a narrow isolated resonance far from a collision threshold is
characterized by a time delay following a familiar Lorentzian profile
\cite{1960-FTS-PR-349,2004-VA-JCP-11675}. In general, at very low
collision energy threshold effects produce deviations from this simple
picture making the analysis less direct \cite{2009-AS-PRA-032701}.  In the
present case the situation is simpler, since the lifetime obtained from
$\mathbf Q^{+}$ shows a pronounced bell shape profile of width much
smaller than its central energy. 

\begin{figure}[htbp]
\includegraphics[width=\columnwidth,clip]{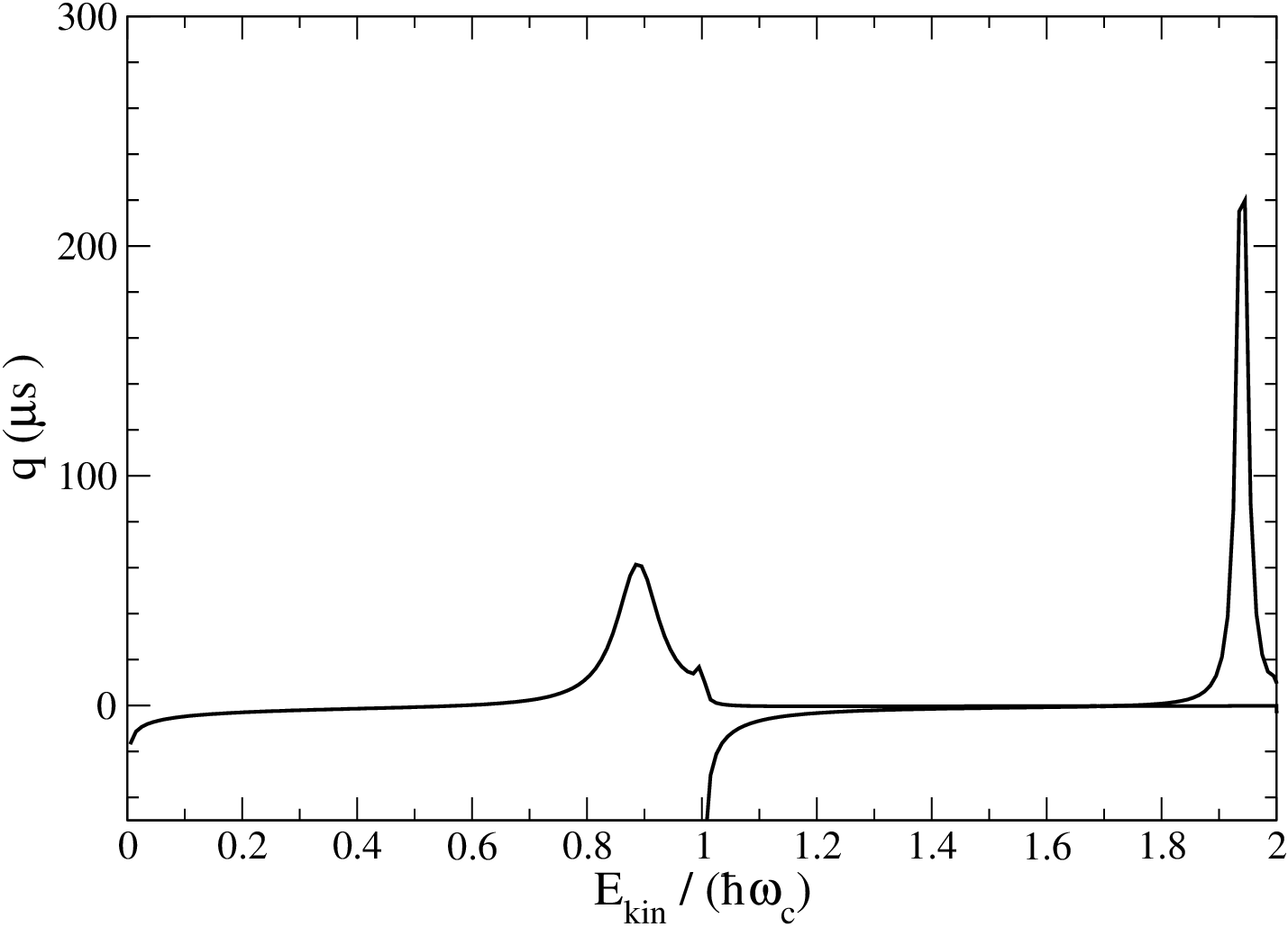}
\caption{The two largest eigenvalues of the time delay matrix as a function of collision energy for the $a_s=R^*$ 
collision in the intermediate field case of Fig.~\ref{fig4}. Only $\pi_z=+$ eigenvalues are shown. 
The long-lived isolated peaks are conclusively of resonant origin.}
\label{fig6}
\end{figure}

This analysis provides conclusive evidence of the resonant nature of the
feature observed in the reflection probability (see Fig.~\ref{fig6}) and
allows us to assign positive parity to the associated metastable state.
As already noted in \cite{2009-AS-PRA-032701} the time delay diverges
positively or negatively at zero energy with a $2 m a_{\rm 1D}/(\hbar
k)$ law, corresponding to the classical time to span the $2a_{\rm 1D}$
distance.

Below the second $n=1$ and third $n=2$ thresholds additional rapid
variations with energy of smaller amplitude appear in $R_{00}$. Again,
the time delay analysis shown in Fig.~\ref{fig6} conclusively confirms the
resonant nature of these features. The dispersive-like shape observed in
the reflection probability arises from interference with background
scattering, whereas the small modulation amplitude is explained by
quenching due to $n=0 \to n^\prime >0$ inelastic processes.
\begin{figure}[htbp]
\includegraphics[width=\columnwidth,clip]{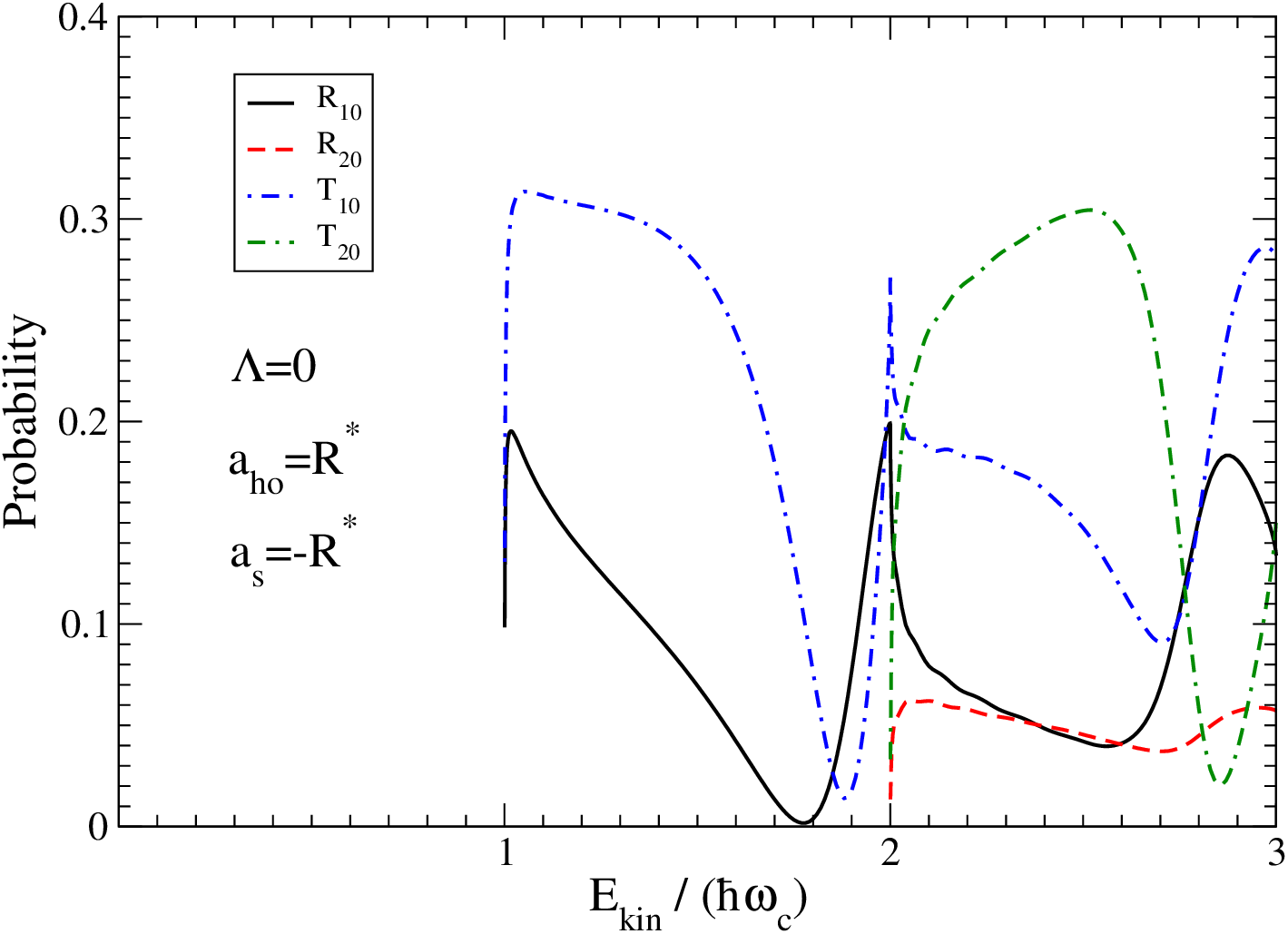}
\caption{Inelastic reflection and transmission probability for an ion in the lowest Landau
level as a function of collision energy in the intermediate field case
$a_{\rm ho}=R^*$ for $\Lambda=0$ and 
$a_s=-R^*$.  
}
\label{fig7}
\end{figure}

Quite strikingly, such resonances seem to occur at similar
locations with respect to each threshold. This property can be
qualitatively interpreted using the pseudopotential analytical method
of Ref.~\cite{2003-TB-PRL-163201}. In this approach the resonance below
the first threshold is assigned to a molecular state generated by the
zero-range interaction in the $n>0$ subspace of closed channels. By a
straightforward generalization consisting in replacing the shift operator
$A^\dagger = \sum_n |\phi_{n+1 , 0} \rangle \langle \phi_{n ,0} |$ below
Eq.~(10) of \cite{2003-TB-PRL-163201} by $A^\dagger = \sum_n |\phi_{n+m ,
0} \rangle \langle \phi_{n , 0} |$, one can show that if a resonant state
is present below the $n=1$ level there has to be a corresponding one
below the $m$th threshold. Similar regularities in the Landau resonance
spectrum have also been proved in low-energy electron photodetachment
using a simplified model which assumes resonant states confined in the
$z=0$ plane \cite{1983-CWC-PRA-83}.

The $a_s=-R^*$ reflection coefficient shows an evident cusp at the
opening of the first threshold. Near the higher thresholds, rounded step
features are also evident. These features arise from the non analyticity
of the scattering amplitude near the opening of a new threshold and do
not have resonant origin. In fact, as shown by Newton both sharp and
rounded features can appear at threshold openings~\cite{1959-JRN-PR-1611}.
\begin{figure}[h!]
\includegraphics[width=\columnwidth,clip]{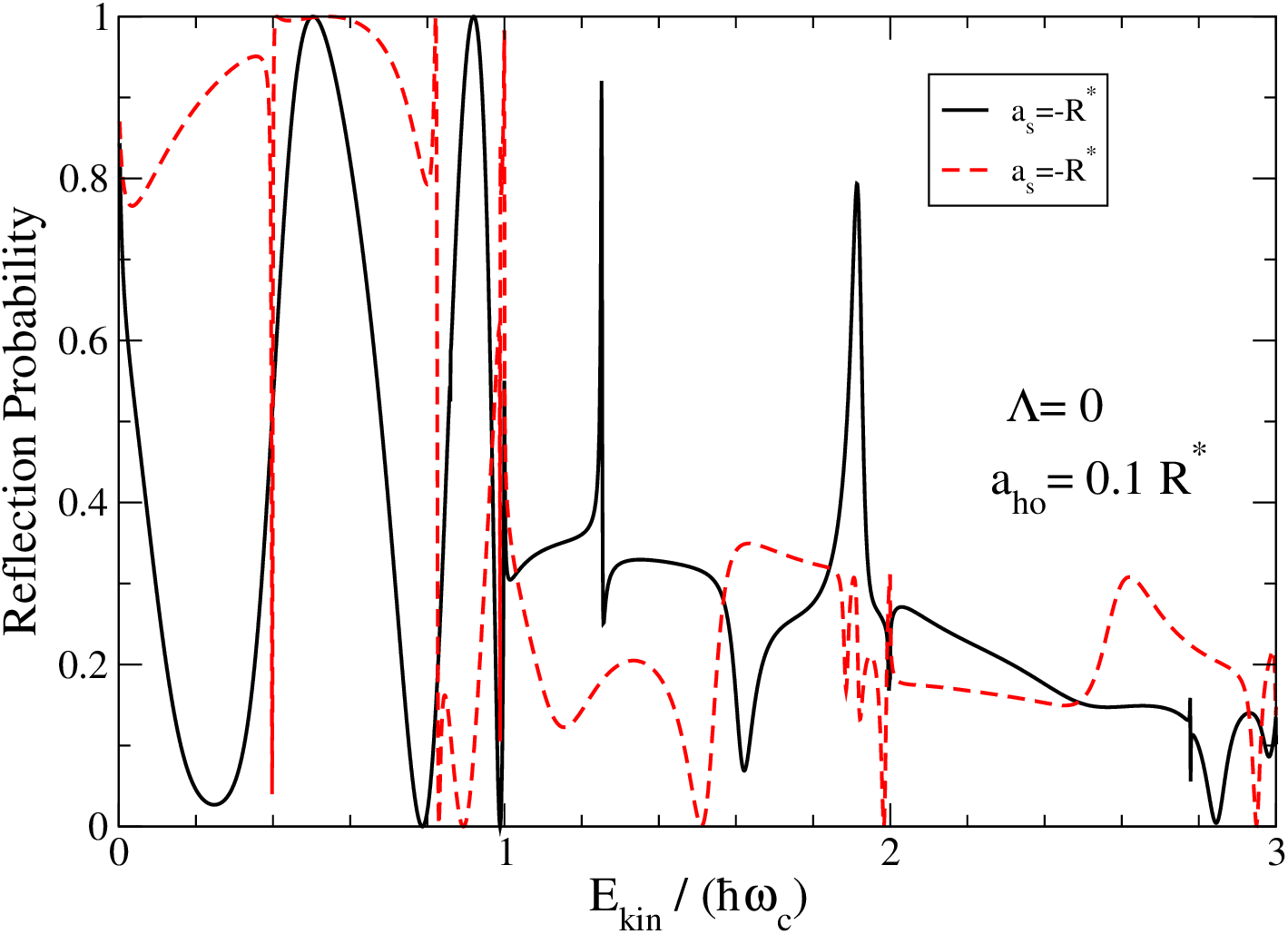}
\caption{
Elastic reflection probability for an ion in the lowest Landau
state as a function of collision energy in the strong field case
$a_{\rm ho}=0.1 R^*$. Axial angular momentum is $\Lambda=0$ and typical
values of the tridimensional scattering length $a_s=\pm R^*$ are selected.
}
\label{fig8}
\end{figure}

We also show in Fig.~\ref{fig7} inelastic transition probabilities from the
lowest to excited Landau states for $a_s=-R^*$. Note that $T_{n0}$ and $R_{n0}$
vanish at the $n$th threshold, in agreement with the Wigner laws, and are
otherwise of significant magnitude. The $T_{n0}$ and $R_{n0}$ differ in
general, showing once again that scattering from odd partial waves is non
negligible in the intermediate field case, and introduces a significant
forward-backward asymmetry. As expected the $R_{n0}$ and $T_{n0}$ do
not vanish at thresholds $m \neq n$, but present cusps and rounded step
singularities similar to the one observed in the elastic probabilities.

Finally, we compute the elastic reflection probability in the strong field
limit (Fig.~\ref{fig8}). The rich resonance structure characteristic of
zero energy scattering has a finite energy counterpart. In fact, multiple
resonance peaks are now observed between Landau thresholds. The narrowest
features are most likely associated to bound states of relatively
large $\ell$ components.  At variance with the intermediate-field case,
a decomposition of the transition matrix identifies the resonance peaks as
originating from both positive and negative parities. Cusp and rounded
steps stemming from the non analytical behaviour of the scattering matrix
and from interference effects between parity components of the scattering
amplitude are also visible near the opening of new energy thresholds.

\section{Conclusions and outlook} 

In conclusion, we have presented a rigorous computational approach
to study atom-ion collisions in the presence of a magnetic field. The
weak-field case maps to a well understood analytical model. In contrast,
we have shown that a novel regime of strong confinement can be reached
in atom-ion systems at experimentally attainable values of the magnetic
field. A series of zero- and finite-energy resonances associated to high order
partial waves appears in this case. The resonance phenomena we have
analyzed in this work should play an important role in applications
based on magnetic control of the atom-ion dynamics.

In a forthcoming paper we will present our results for the qualitatively
different case of $|\Lambda| >0$ collisions and consider the case of a
scattering center not necessarily placed on the axis of the cyclotron
orbit. In perspective, it will be interesting to investigate the
interplay of pre-existing atom-ion magnetic Feshbach resonances and of
Landau quantization. Moreover, in cases where the atom cannot be treated
as infinitely massive, quasi separability of the center of mass motion
might give rise to still unexplored resonance phenomena.

\section{Acknowledgments}
A. S. gratefully acknowledges support from CNRS and Rennes-Metropole.

% \bibliographystyle{unsrt}
% \bibliographystyle{prsty}
% \bibliographystyle{natbib}
%\bibliography{Journal,Biblio}

\section{References}


\begin{thebibliography}{10}

\bibitem{2000-HW-PRA-052704}
H.~Wang, A.~N. Nikolov, J.~R. Ensher, P.~L. Gould, E.~E. Eyler, W.~C. Stwalley,
  J.~P. Burke, J.~L. Bohn, Chris.~H. Greene, E.~Tiesinga, C.~J. Williams, and
  P.~S. Julienne.
\newblock Ground-state scattering lengths for potassium isotopes determined by
  double-resonance photoassociative spectroscopy of ultracold $^{39}${K}.
\newblock {\em Phys. Rev. A}, 62:052704--1--4, 2000.

\bibitem{2008-AS-PRA-052705}
A.~Simoni, M.~Zaccanti, C.~D'Errico, M.~Fattori, G.~Roati, M.~Inguscio, and
  G.~Modugno.
\newblock Near-threshold model for ultracold {K}{R}b dimers from interisotope
  {F}eshbach spectroscopy.
\newblock {\em Phys. Rev. A}, 77:052705--1--8, 2008.

\bibitem{2004-CAR-PRL-040403}
C.~A. Regal, M.~Greiner, and D.~S. Jin.
\newblock Observation of resonance condensation of fermionic atom pairs.
\newblock {\em Phys. Rev. Lett.}, 92:040403--1--4, 2004.

\bibitem{2006-TV-NP-692}
T.~Volz, N.~Syassen, D.~M. Bauer, E.~Hansis, S.~D{\"u}rr, and G.~Rempe.
\newblock Preparation of a quantum state with one molecule at each site of an
  optical lattice.
\newblock {\em Nature Physics}, 2:692--695, 2006.

\bibitem{1981-JBD-RMP-287}
J.~B. Delos.
\newblock Theory of electronic transitions in slow atomic collisions.
\newblock {\em Rev. Mod. Phys}, 53:287--357, 1981.

\bibitem{2010-HD-PRA-012708}
H.~Doerk, Z.~Idziaszek, and T.~Calarco.
\newblock Atom-ion quantum gate.
\newblock {\em Phys. Rev. A}, 81:012708--1--13, 2010.

\bibitem{2002-RC-PRL-093001}
R.~C\^ot\'e, V.~Kharchenko, and M.~D. Lukin.
\newblock Mesoscopic molecular ions in {B}ose-{E}instein condensates.
\newblock {\em Phys. Rev. Lett.}, 89:093001--1--4, 2002.

\bibitem{2010-CZ-NAT-388}
C.~Zipkes, S.~Palzer, C.~Sias, and M.~K{\"o}hl.
\newblock A trapped single ion inside a {B}ose-{E}instein condensate.
\newblock {\em Nature}, 464:388--391, 2010.

\bibitem{2009-ATG-PRL-223201}
A.~T. Grier, M.~Cetina, F.~Oru\v{c}evi\'{c}, and V. Vuleti\'{c}.
\newblock Observation of cold collisions between trapped ions and trapped
  atoms.
\newblock {\em Phys. Rev. Lett.}, 102:223201--1--4, 2009.

\bibitem{2009-ZI-PRA-010702}
Z.~Idziaszek, T.~Calarco, P.~S. Julienne, and A.~Simoni.
\newblock Quantum theory of ultracold atom-ion collisions.
\newblock {\em Phys. Rev. A}, 79:010702(R)--1--4, 2009.

\bibitem{2003-OPM-PRA-042705}
O.~P. Makarov, R.~C\^ot\'e, H.~Michels, and W.~W. Smith.
\newblock Radiative charge-transfer lifetime of the excited state of ${\rm
  (naca)}^+$.
\newblock {\em Phys. Rev. A}, 67:042705--1--5, 2003.

\bibitem{2010-JG-PRA-041601}
J.~Goold, H.~Doerk, Z.~Idziaszek, T.~Calarco, and Th. Busch.
\newblock Ion-induced density bubble in a strongly correlated one-dimensional
  gas.
\newblock {\em Phys. Rev. A}, 81:041601(R)--1--4, 2010.

\bibitem{2007-ZI-PRA-033409}
Z.~Idziaszek, T.~Calarco, and P.~Zoller.
\newblock Controlled collisions of a single atom and an ion guided by movable
  trapping potentials.
\newblock {\em Phys. Rev. A}, 76:033409--1--16, 2007.

\bibitem{BK-1999-LDL}
L.~D. Landau and E.~M. Lifschitz.
\newblock Quantum mechanics.
\newblock {\em (Butterworth-Heinemann, Oxford)}, 1999.

\bibitem{1978-WAMB-PRL-1320}
W.~A.~M. Blumberg, R.~M. Jopson, and D.~J. Larson.
\newblock Precision laser photodetachment spectroscopy in magnetic fields.
\newblock {\em Phys. Rev. Lett.}, 40:1320--1323, 1978.

\bibitem{1987-CHG-PRA-4236}
C-H. Greene.
\newblock Negative-ion photodetachment in a weak magnetic field.
\newblock {\em Phys. Rev. A}, 36:4236--4244, 1987.

\bibitem{1991-QW-PRA-7448}
Q.~Wang and C.~H. Greene.
\newblock $r$-matrix calculation of atomic hydrogen photoionization in a strong
  magnetic field.
\newblock {\em Phys. Rev. A}, 44:7448--7458, 1991.

\bibitem{1977-JEV-ANNP-431}
J.~E. Avron, I.~W. Herbst, and B.~Simon.
\newblock Separation of center of mass in homogeneous magnetic fields.
\newblock {\em Ann. Phys.}, 114:431--451, 1977.

\bibitem{1996-ET-JRNIST-505}
E.~Tiesinga, C.~J. Williams, P.~S. Julienne, K.~M. Jones, P.~D. Lett, and W.~D.
  Phillips.
\newblock A spectroscopic determination of scattering lengths for sodium atom
  collisions.
\newblock {\em J. Res. Natl. Inst. Stand. Tech.}, 101:505--520, 1996.

\bibitem{1989-PSJ-JOSAB-2257}
P.~S. Julienne and F.~H. Mies.
\newblock Collisions of ultracold trapped atoms.
\newblock {\em J. Opt. Soc. Am. B}, 6:2257--2269, 1989.

\bibitem{BK-1964-MLG}
M.~L. Goldberger and K.~M. Watson.
\newblock Collision theory.
\newblock {\em (Wiley, New York)}, 1964.

\bibitem{1960-LMD-NUP-275}
L.~M. Delves.
\newblock Tertiary and general-order collisions (ii).
\newblock {\em Nucl. Phys.}, 20:275--308, 1960.

\bibitem{1998-MO-PRL-938}
M.~Olshanii.
\newblock Atomic scattering in the presence of an external confinement and a
  gas of impenetrable bosons.
\newblock {\em Phys. Rev. Lett.}, 81:938--941, 1998.

\bibitem{2004-BP-NAT-277}
B.~Paredes, A.~Widera, V.~Murg, O.~Mandel, S.~Folling, I.~Cirac, G.~V.
  Shlyapnikov, T.~W. Hansch, and I.~Bloch.
\newblock {T}onks-{G}irardeau gas of ultracold atoms in an optical lattice.
\newblock {\em Nature}, 429:277--281, 2004.

\bibitem{1958-HF-ANNP-357}
H.~Feshbach.
\newblock Unified theory of nuclear reactions.
\newblock {\em Ann. Phys.}, 5:357--390, 1958.

\bibitem{1961-UF-PRA-1866}
U.~Fano.
\newblock Effects of configuration interaction on intensities and phase shifts.
\newblock {\em Phys. Rev. A}, 124:1866--1878, 1961.

\bibitem{1986-DEM-JCP-6425}
D.~E. Manolopoulos.
\newblock An improved log derivative method for inelastic scattering.
\newblock {\em J. Chem. Phys.}, 85:6425--6429, 1986.

\bibitem{1989-JML-CPL-178}
J.-M. Launay and M.~{Le Dourneuf}.
\newblock Hyperspherical close-coupling calculation of integral cross sections
  for the reaction {H}+{H}$_2$ $\to$ {H}$_2$+h.
\newblock {\em Chem. Phys. Letters}, 163:178--187, 1989.

\bibitem{1987-RTP-JCP-3888}
R.~T. Pack and G.~A. Parker.
\newblock Quantum reactive scattering in three dimensions using
  hyperspherical(aph) coordinates. theory.
\newblock {\em J. Chem. Phys.}, 87:3888--3921, 1987.

\bibitem{1995-TPG-PRA-607}
T.~P. Grozdanov.
\newblock Photodetachment of an electron bound by a zero-range potential in
  magnetic fields of arbitrary strength.
\newblock {\em Phys. Rev. A}, 51:607--610, 1995.

\bibitem{2002-ELB-PRA-013403}
E.~L. Bolda, E.~Tiesinga, and P.~S. Julienne.
\newblock Effective-scattering-length model of ultracold atomic collisions and
  feshbach resonances in tight harmonic traps.
\newblock {\em Phys. Rev. A}, 66:013403--1--7, 2002.

\bibitem{2006-JIK-PRL-193203}
J.~I. Kim, V.~S. Melezhik, and P.~Schmelcher.
\newblock Suppression of quantum scattering in strongly confined systems.
\newblock {\em Phys. Rev. Lett.}, 97:193203--1--4, 2006.

\bibitem{1960-FTS-PR-349}
F.~T. Smith.
\newblock Lifetime matrix in collision theory.
\newblock {\em Phys. Rep.}, 118:349--356, 1960.

\bibitem{2004-VA-JCP-11675}
V.~Aquilanti, S.~Cavalli, A.~Simoni, A.~Aguilar, J.~M. Lucas, and D.~De Fazio.
\newblock Lifetime of reactive scattering resonances: $q$-matrix analysis and
  angular momentum dependence for the {F}+{H}$_2$ reaction by the
  hyperquantization algorithm.
\newblock {\em J. Chem. Phys.}, 121:11675--11690, 2004.

\bibitem{2009-AS-PRA-032701}
A.~Simoni, J-M. Launay, and P.~Sold\'an.
\newblock {F}eshbach resonances in ultracold atom-molecule collisions.
\newblock {\em Phys. Rev. A}, 79:032701--1--6, 2009.

\bibitem{2003-TB-PRL-163201}
T.~Bergeman, M.~G. Moore, and M.~Olshanii.
\newblock Atom-atom scattering under cylindrical harmonic confinement:
  Numerical and analytic studies of the confinement induced resonance.
\newblock {\em Phys. Rev. Lett.}, 91:163201--1--4, 2003.

\bibitem{1983-CWC-PRA-83}
C.~W. Clark.
\newblock Low-energy electron-atom scattering in a magnetic field.
\newblock {\em Phys. Rev. A}, 28:83--90, 1983.

\bibitem{1959-JRN-PR-1611}
R.~G. Newton.
\newblock Threshold properties of scattering and reaction cross sections.
\newblock {\em Phys. Rep.}, 114:1611--1618, 1959.

\end{thebibliography}
\end{document}